\documentclass[%
 reprint,
superscriptaddress,
showpacs,
preprintnumbers,
 amsmath,amssymb,
prx,
longbibliography
]{revtex4-1}

\usepackage{graphicx}
\usepackage{dcolumn}
\usepackage{bm}
\usepackage{setspace}	
\usepackage{color}
\usepackage{amsmath}	
\usepackage{upgreek}

\DeclareGraphicsExtensions{.eps}

\def\degree{\mbox{$^\circ$}}

\def\ii{{\mathrm{i}}}
\def\ff{{\mathrm{f}}}
\def\ee{{\mathrm{e}}}
\def\dd{{\mathrm{d}}}

\def\HH{{\mathrm{H}}}
\def\VV{{\mathrm{V}}}
\def\DD{{\mathrm{D}}}
\def\AA{{\mathrm{A}}}

\def\cc{\mathrm{c.c.}} 
\def\Hc{\mathrm{H.c.}} 

\def\no{{\nonumber}} 

\def\bra#1{\langle #1|}

\def\ket#1{|#1\rangle}

\def\bracket#1{\langle #1 \rangle}
\def\bracketi#1#2{\langle #1 | #2 \rangle}
\def\bracketii#1#2#3{\langle #1 | #2| #3\rangle}

\def\sub#1{_\mathrm{#1}} 
\def\sur#1{^\mathrm{#1}} 

\def\tr{\mathrm{tr}}

\begin{document}


\title{
Complex counterpart of variance in quantum measurements\\
for pre- and post-selected systems
}


\author{Kazuhisa Ogawa}
\email{ogawak@ist.hokudai.ac.jp}
\affiliation{%
Graduate School of Information Science and Technology, Hokkaido University, Sapporo 060-0814, Japan
}%
 
\author{Natsuki Abe}
\affiliation{%
Graduate School of Information Science and Technology, Hokkaido University, Sapporo 060-0814, Japan
}%

\author{Hirokazu Kobayashi}
\affiliation{%
School of System Engineering, Kochi University of Technology, Tosayamada-cho, Kochi 782-8502, Japan
}%

\author{Akihisa Tomita}
\affiliation{%
Graduate School of Information Science and Technology, Hokkaido University, Sapporo 060-0814, Japan
}%

\date{\today}

\begin{abstract}
The variance of an observable in a pre-selected quantum system, which is always real and non-negative, appears as an increase in the probe wave packet width in indirect measurements.
Extending this framework to pre- and post-selected systems, we formulate a complex-valued counterpart of the variance called ``weak variance.''
In our formulation, the real and imaginary parts of the weak variance appear as changes in the probe wave packet width in the vertical--horizontal and diagonal--antidiagonal directions, respectively, on the quadrature phase plane.
Using an optical system, we experimentally demonstrate these changes in the probe wave packet width caused by the real negative and purely imaginary weak variances.
Furthermore, we show that the weak variance can be expressed as the variance of the weak-valued probability distribution in pre- and post-selected systems.
These operational and statistical interpretations support the rationality of formulating the weak variance as a complex counterpart of the variance in pre- and post-selected systems.
\end{abstract}


\maketitle


\begin{spacing}{1.2}

\section{Introduction}

The outcomes of quantum measurements show probabilistic behavior.
This characteristic, which is not observed in classical systems, has been the root of many fundamental arguments in quantum theory \cite{wheeler2014quantum}.
In the quantum measurement of an observable $\hat{A}$, the probabilistic behavior of its measurement outcomes is characterized by measurement statistics such as expectation value $\bracket{\hat{A}}$ and variance $\sigma^2(\hat{A})$.
These values are generally measured using an indirect measurement method \cite{neumann}.
In indirect measurement, the target system to be measured is coupled with an external probe system through von Neumann interaction.
Regardless of the coupling strength, the expectation value $\bracket{\hat{A}}$ and variance $\sigma^2(\hat{A})$ in the target system are obtained from the displacement of the probe wave packet and the increase in its width, respectively.
In other words, the probe wave packet in an indirect measurement serves as the interface that displays the probabilistic characteristics of the target system.

Interestingly, when the target system is further post-selected, the displacement of the probe wave packet differs from
$\bracket{\hat{A}}$.
In particular, when the coupling strength is weak (weak measurement setup), the probe displacement is given by $\mathrm{Re}\bracket{\hat{A}}\sub{w}$, where $\bracket{\hat{A}}\sub{w}:=\bracketii{\ff}{\hat{A}}{\ii}/\bracketi{\ff}{\ii}$ in the pre- and post-selected system $\{\ket{\ii},\ket{\ff}\}$ is called weak value \cite{PhysRevLett.60.1351}.
$\bracket{\hat{A}}\sub{w}$ is complex in general and can exceed the spectral range of $\hat{A}$.
By regarding the weak value as a complex counterpart of the expectation value in the pre- and post-selected system,
new approaches to fundamental problems in quantum mechanics involving pre- and post-selection have been investigated, such as various quantum paradoxes \cite{aharonov1991complete,resch2004experimental,PhysRevLett.102.020404,yokota2009direct,denkmayr2014observation,okamoto2016experimental,PhysRevLett.111.240402}, understanding of the violation of Bell's inequality using negative probabilities \cite{PhysRevA.91.012113}, the relationship between disturbance and complementarity in quantum measurements \cite{wiseman2003directly,mir2007double,xiao2019observing}, verification of the uncertainty relations \cite{PhysRevLett.112.020401,PhysRevLett.112.020402,PhysRevLett.109.100404}, observation of Bohmian trajectories \cite{kocsis2011observing,mahler2016experimental}, and demonstration of the violation of macrorealism \cite{PhysRevLett.106.040402,goggin2011violation}.

%


Similar to the relation between the weak value and expectation value, does there exist a counterpart of the variance in pre- and post-selected systems?
To answer this question, we consider the function of the probe wave packet in indirect measurement as an interface that displays the characteristics of the target system. 
As mentioned earlier, the variance $\sigma^2(\hat{A})$ in a pre-selected system manifests as an increase in the probe wave packet width in indirect measurement, and owing to the non-negativity of the variance, the wave packet width never decreases.
However, for pre- and post-selected systems, any counterparts of the variance cannot be observed in the typical framework of the weak measurement \cite{PhysRevLett.60.1351}, in which the probe wave packet width does not change because the second- and higher-order terms of the coupling strength are ignored.
Here, we focus on the recent studies reporting that when considering the second- and higher-order terms of the coupling strength, the probe wave packet width can not only increase but also decrease under appropriate pre- and post-selection conditions \cite{de2015uncertainty,matsuoka2017generation}.
If these reported phenomena are interpreted to result from a counterpart of the variance in pre- and post-selected systems, it may be possible to formulate an effective variance-like quantity that can be negative.

In this study, we investigate the general changes in the width of the probe wave packet during indirect measurements of pre- and post-selected systems. We then formulate a counterpart of the variance in these systems. 
This counterpart, denoted here as \textit{weak variance}, can indeed be negative and manifests as the decrease in the probe wave packet width.
Moreover, the weak variance is generally complex and can be understood by observing the changes in the probe wave packet width on the quadrature phase plane. 
To demonstrate this phenomenon, we conducted an optical experiment for observing the changes in the beam packet width in proportion to the real and imaginary parts of the complex weak variance.
In addition, to clarify the concept of weak variance, we express the weak variance as the second-order moment of the weak-valued probability distribution \cite{aharonov1991complete,resch2004experimental,PhysRevLett.102.020404,yokota2009direct,denkmayr2014observation,okamoto2016experimental,PhysRevLett.111.240402,wiseman2003directly,mir2007double,PhysRevA.91.012113,arvidsson1903quantum,PhysRevA.97.042105,PhysRevLett.122.040404}, which is a quasi-probability distribution in pre- and post-selected systems.
Based on the agreement between the operational and statistical interpretations, we propose that our weak variance can be considered a reasonable definition of a complex counterpart of the variance in pre- and post-selected systems than previous formulations \cite{PhysRevA.52.2538,reznik1995interaction,tanaka2002semiclassical,brodutch2008weak,parks2014weak,PhysRevA.41.11,parks2018weak,feyereisen2015weak,hofmann2011characterization,hofmann2011role,pati2014uncertainty,song2015uncertainty,hofer2017quasi}.
Furthermore, we formulate a counterpart of the higher-order moment and investigate its operational and statistical meanings and applications.

\section{Weak variance appearing in indirect measurement for pre- and post-selected systems}

\begin{figure}
\includegraphics[width=8.5cm]{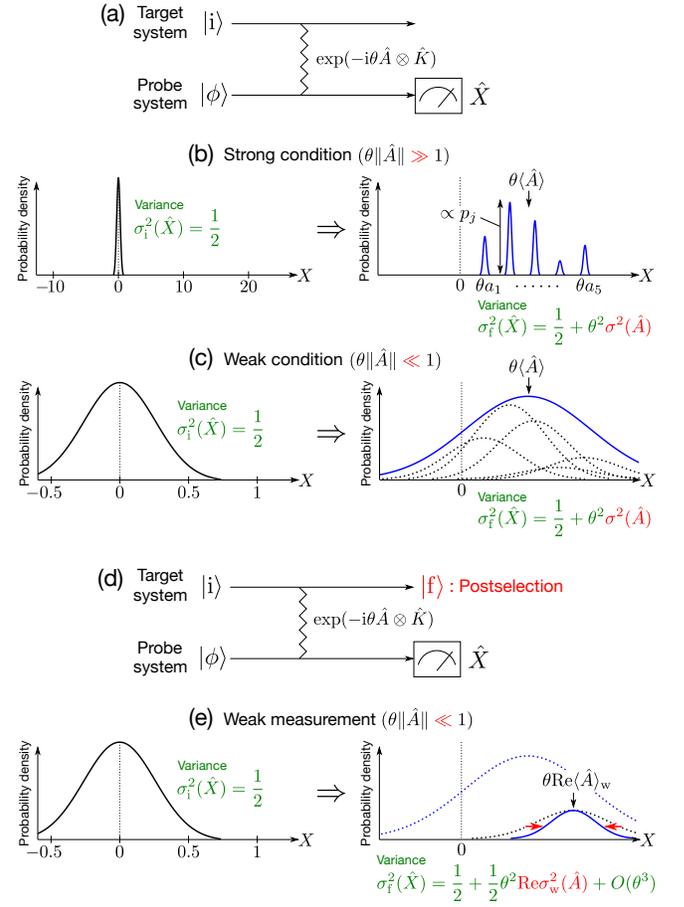}
\caption{
(a) Quantum circuit of indirect measurements of the pre-selected system $\ket{\ii}$.
(b) Change in the probe wave packet caused by interactions in the quantum circuit (a) under the strong coupling condition ($\theta\|\hat{A}\|\gg 1$).
The distribution of probe wave packets after the interaction reproduces the probability distribution of the outcomes of projective measurements of $\hat{A}$ in $\ket{\ii}$. 
 (c) Change in the probe wave packet under the weak coupling condition ($\theta\|\hat{A}\|\ll 1$), where the horizontal axis has been rescaled from that of (b).
The variance of the probe wave packet after the interaction increases in proportion to the variance $\sigma^2(\hat{A})$ and never decreases.
 (d) Quantum circuit of indirect measurements of the pre- and post-selected system $\{\ket{\ii}, \ket{\ff}\}$ (weak measurement setup).
(e) Change in the probe wave packet in the weak measurement circuit (d). 
The real part of the weak variance appears in the variance change of the probe wave packet after the post-selection.
Unlike the pre-selected system (c), the pre- and post-selected system admits a narrowed variance of the probe wave packet when the real part of the weak variance becomes negative.
}\label{fig:theory}
\end{figure}

The indirect measurements have been made with a Gaussian probe as shown in Figs.~\ref{fig:theory}(a) and (d). 
After reviewing these measurements, we explain the appearance of a complex weak variance in the pre- and post-selected system.
The target system to be measured and the probe system are pre-selected in states $\ket{\ii}$ and $\ket{\phi}$, respectively.
The initial probe state $\ket{\phi}$ can be expanded as $\ket{\phi}=\int_{-\infty}^\infty\dd X\phi(X)\ket{X}$, where the wave function $\phi(X)$ is the Gaussian distribution $\phi(X)=\pi^{-1/4}\exp(-{X^2}/{2})$ and $X$ is a dimensionless variable \footnote{The dimensionless variable $X$ is obtained by dividing the position variable $x$ by the standard deviation $\sigma$ of the wave function $\pi^{-1/4}\sigma^{-1/2}\exp[-x^2/(2\sigma^2)]$.}.
The observable of the dimensionless position $\hat{X}$ can be spectrally decomposed as $\hat{X}=\int_{-\infty}^\infty \dd X X \ket{X}\bra{X}$.
The time evolution by the interaction Hamiltonian $\hat{A}\otimes\hat{K}$ is represented by the unitary operator $\hat{U}(\theta)=\exp(-\ii\theta\hat{A}\otimes\hat{K})$, where $\hat{A}=\sum_ja_j\hat{\varPi}_j$ is the observable to be measured in the target system, $a_j$ is an eigenvalue of $\hat{A}$, $\hat{\varPi}_j$ is the projector onto the eigenspace of $\hat{A}$ belonging to eigenvalue $a_j$, $\hat{K}$ is the canonical conjugate observable of $\hat{X}$ satisfying $[\hat{X},\hat{K}]=\ii\hat{1}$, and $\theta$ is a parameter with the reciprocal dimension of $\hat{A}$.
The coupling strength is characterized by $\theta\|\hat{A}\|$, where $\|\hat{A}\|$ is the largest eigenvalue of $\hat{A}$: if $\theta\|\hat{A}\|\gg 1$ ($\ll 1$), the coupling is considered strong (weak).

Let us consider an indirect measurement of the observable $\hat{A}$, as shown in Fig.~\ref{fig:theory} (a).
Suppose that measurement $\hat{X}$ in the probe system is made to the state after the interaction, $\ket{\varPsi}:=\exp(-\ii\theta\hat{A}\otimes\hat{K})\ket{\ii}\ket{\phi}$.
The probability distribution $P(X)$ of obtaining the result $X$ is 
\begin{align}
P(X)=|\bracketi{X}{\varPsi}|^2=\sum_jp_j|\phi(X-\theta a_j)|^2,
\end{align}
where $p_j:=\bracketii{\ii}{\hat{\varPi}_j}{\ii}$ is the projection probability of $\ket{\ii}$ onto $\hat{\varPi}_j$. 
If the coupling is strong ($\theta\|\hat{A}\|\gg 1 $), the wave packet $|\phi(X-\theta a_j)|^2$ for each $j$ is well separated from other wave packets, and $P(X)$ reproduces the probability distribution $\{p_j\}_j$ [Fig.~\ref{fig:theory}(b)].
However, if the coupling is weak ($\theta\|\hat{A}\|\ll 1 $), the wave packets overlap and $P(X)$ does not reproduce $\{p_j\}_j$ [Fig.~\ref{fig:theory}(c)].
Nevertheless, regardless of the coupling strength, the statistics of $\hat{A}$ in the target system $\ket{\ii}$, such as the expectation value $\bracket{\hat{A}}$ and the variance $\sigma^2(\hat{A})$, can be acquired from the changes in the probe distribution $P(X)$.
The expectation value and variance of $X$ in $P(X)$ are respectively expressed as
\begin{align}
\bracket{\hat{X}}\sub{f}=\bracket{\hat{X}}\sub{i}+\theta\bracket{\hat{A}}, \quad
\sigma^2\sub{f}(\hat{X})=\sigma^2\sub{i}(\hat{X})+\theta^2\sigma^2(\hat{A}),\label{eq:4}
\end{align}
where $\bracket{\hat{X}}\sub{i}$ and $\sigma^2\sub{i}(\hat{X})$ are the expectation value and variance of $\hat{X}$ in the initial probe state $\ket{\phi}$, respectively. In this case, $\bracket{\hat{X}}\sub{i}=0$ and $\sigma^2\sub{i}(\hat{X})=1/2$.
Therefore, the expectation value $\bracket{\hat{A}}$ and variance $\sigma^2(\hat{A})$ can be measured under both strong and weak coupling conditions.
Here, we stress that after the interaction, the variance of the probe wave packet $\sigma^2\sub{f}(\hat{X})$ never decreases because the variance $\sigma^2(\hat{A})$ is non-negative.

We next consider that the target system is pre- and post-selected in states $\ket{\ii}$ and $\ket{\ff}$, respectively [Fig.~\ref{fig:theory}(d)].
The non-normalized state of the probe system after the post-selection $\ket{\tilde{\phi}\sub{f}}:=\bracketi{\ff}{\varPsi}$ is represented as
\begin{align}
\ket{\tilde{\phi}\sub{f}}
=\bracketi{\ff}{\ii}
\left(\hat{1}-\ii\theta\bracket{\hat{A}}\sub{w}\hat{K}
-\frac{\theta^2}{2}\bracket{\hat{A}^2}\sub{w}\hat{K}^2\right)
\ket{\phi}
+O(\theta^3).\label{eq:5}
\end{align}
The expectation value of $\hat{X}$ in the non-normalized state $\ket{\tilde{\phi}\sub{f}}$ is $\bracket{\hat{X}}\sub{f}={\bracketii{\tilde{\phi}\sub{f}}{\hat{X}}{\tilde{\phi}\sub{f}}}/{\bracketi{\tilde{\phi}\sub{f}}{\tilde{\phi}\sub{f}}}=\mathrm{Re}\bracket{\hat{A}}\sub{w}\theta+O(\theta^3)$. The real part of the weak value $\bracket{\hat{A}}\sub{w}=\bracketii{\ff}{\hat{A}}{\ii}/\bracketi{\ff}{\ii}$ appears in the displacement of the probe wave packet, as previously reported for weak measurements \cite{PhysRevLett.60.1351}.
The imaginary part of the weak value is observed in the displacement of the probe wave packet in the $\hat{K}$ basis: $\bracket{\hat{K}}\sub{f}=\mathrm{Im}\bracket{\hat{A}}\sub{w}\theta+O(\theta^3)$ \cite{jozsa2007complex}.
By introducing the generalized position operator $\hat{M}:=\hat{X}\cos\alpha+\hat{K}\sin\alpha$ ($\alpha\in[0,2\pi)$), these relations can be summarized as 
\begin{align}
 \bracket{\hat{M}}\sub{f}
&=\left(\cos\alpha\,\mathrm{Re}\bracket{\hat{A}}\sub{w}+\sin\alpha\,\mathrm{Im}\bracket{\hat{A}}\sub{w}\right)\theta+O(\theta^3).\label{eq:1}
\end{align}
%

Now let us examine the change in the probe wave packet width. 
The variance of $\hat{X}$ for $\ket{\tilde{\phi}\sub{f}}$ is calculated as
\begin{align}
 \sigma^2\sub{f}(\hat{X})&=\bracket{\hat{X}^2}\sub{f}-\bracket{\hat{X}}\sub{f}^2\no\\
&=\sigma\sub{i}^2(\hat{X})
+\frac{1}{2}\mathrm{Re}
\left(\bracket{\hat{A}^2}\sub{w}-\bracket{\hat{A}}\sub{w}^2\right)
\theta^2+O(\theta^3).
\end{align}
The real part of the variance-like quantity appears in the quadratic term of $\theta$, which is ignored in the conventional weak measurement context. 
We define this quantity as the \textit{weak variance} $\sigma\sub{w}^2(\hat{A})$ of $\hat{A}$:
\begin{align}
 \sigma\sub{w}^2(\hat{A})
&:=\bracket{\hat{A}^2}\sub{w}-\bracket{\hat{A}}\sub{w}^2.\label{eq:3}
\end{align}
The real part of the weak variance is similar to normal variance in that it appears as a change in the probe wave packet width in the $\hat{X}$ basis [Eq.~(\ref{eq:4})]. 
However, unlike the normal variance, the weak variance can be negative, in which case the wave packet width then decreases as shown in Fig.~\ref{fig:theory}(e).
The decrease in the probe wave packet width reported in previous studies \cite{de2015uncertainty,matsuoka2017generation} can be reinterpreted as the effect of the negative weak variance.

\begin{figure}
\includegraphics[width=8.5cm]{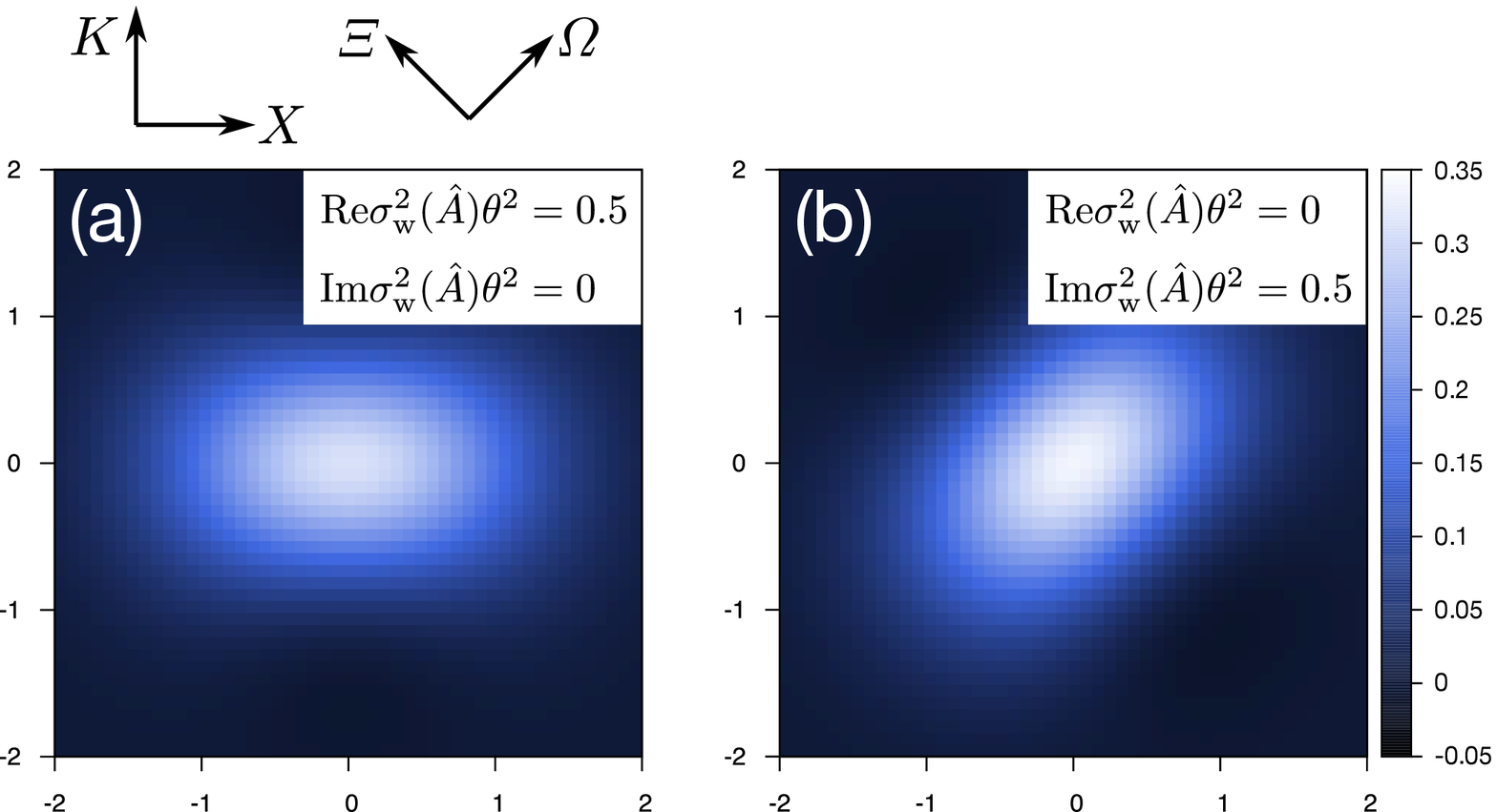}
\caption{
Wigner functions of the normalized probe states after post-selection, $\ket{\tilde{\phi}\sub{f}}/\|\ket{\tilde{\phi}\sub{f}}\|$, assuming $\bracket{\hat{A}}\sub{w}=0$ and neglecting $O(\theta^3)$ in Eq.~(\ref{eq:5}).
The horizontal and vertical axes represent the observables $\hat{X}$ and $\hat{K}$, respectively, and the 45$\degree$ and 135$\degree$ axes represent the observables $\hat{\varOmega}$ and $\hat{\varXi}$, respectively.
(a) When $\mathrm{Re}\sigma\sub{w}^2(\hat{A})\theta^2=0.5$ and $\mathrm{Im}\sigma\sub{w}^2(\hat{A})\theta^2=0$, the wave packet spreads along the $X$ axis and narrows along the $K$ axis.
(b) When $\mathrm{Re}\sigma\sub{w}^2(\hat{A})\theta^2=0$ and $\mathrm{Im}\sigma\sub{w}^2(\hat{A})\theta^2=0.5$, the wave packet spreads along the $\varOmega$ axis and narrows along the $\varXi$ axis.
 }\label{fig:theory2}
\end{figure}

We next consider the appearance of the imaginary part of the weak variance.
The variance of the generalized position operator $\hat{M}$ for $\ket{\tilde{\phi}\sub{f}}$ is calculated as (see Appendix~\ref{sec:appA} for a detailed analysis of mixed pre- and post-selected states of the target system)
\begin{align}
 \sigma^2\sub{f}(\hat{M})
&=\frac{1}{2}
+\frac{1}{2}\Big[
\cos(2\alpha)\mathrm{Re}\sigma^2\sub{w}(\hat{A})\no\\
&\hspace{1.5cm}+\sin(2\alpha)\mathrm{Im}\sigma^2\sub{w}(\hat{A})
\Big]\theta^2+O(\theta^3).\label{eq:2}
\end{align}
This equation indicates that the real and imaginary parts of the weak variance appear in the changes in the probe wave packet width in different measurement bases. 
For example, when choosing $\hat{M}=\hat{K}$ ($\alpha=\pi/2$), the variance of $\hat{K}$ in state $\ket{\tilde{\phi}\sub{f}}$ is given as $\sigma\sub{f}^2(\hat{K})=\sigma\sub{i}^2(\hat{K})-(1/2)\mathrm{Re}\sigma^2\sub{w}(\hat{A})\theta^2+O(\theta^3)$; that is, $\mathrm{Re}\sigma^2\sub{w}(\hat{A})$ also appears in the width change of the wave packet in the $\hat{K}$ basis.
To clarify these relations, we plot them on the quadrature phase plane [Fig.~\ref{fig:theory2}(a)].
When $\mathrm{Re}\sigma\sub{w}^2(\hat{A})>0$, the wave packet spreads along the $X$ (horizontal) axis, while it narrows along the $K$ (vertical) axis.
This relationship satisfies the Kennard--Robertson uncertainty relation \cite{kennerd,PhysRev.34.163} up to the quadratic of $\theta$: $\sigma\sub{f}^2(\hat{X})\sigma\sub{f}^2(\hat{K})=1/4+O(\theta^3)$.

However, when $\alpha=\pi/4$, the measured observable becomes $\hat{M}=(\hat{X}+\hat{K})/\sqrt{2}=:\hat{\varOmega}$, which corresponds to the 45$\degree$ axis in the quadrature phase plane of Fig.~\ref{fig:theory2}.
Through the relation $\sigma\sub{f}^2(\hat{\varOmega})=\sigma\sub{i}^2(\hat{\varOmega})+(1/2)\mathrm{Im}\sigma^2\sub{w}(\hat{A})\theta^2+O(\theta^3)$, the imaginary part of the weak variance $\mathrm{Im}\sigma^2\sub{w}(\hat{A})$ can be observed as the change in the probe wave packet width in the $\hat{\varOmega}$ basis. 
Moreover, when $\alpha=3\pi/4$, the measured observable becomes $\hat{M}=(-\hat{X}+\hat{K})/\sqrt{2}=:\hat{\varXi}$, which is the canonical conjugate of $\hat{\varOmega}$. This observable satisfies $[\hat{\varOmega},\hat{\varXi}]=\ii\hat{1}$ and corresponds to the 135$\degree$ axis in the quadrature phase plane of Fig.~\ref{fig:theory2}.
Through the relation $\sigma\sub{f}^2(\hat{\varXi})=\sigma\sub{i}^2(\hat{\varXi})-(1/2)\mathrm{Im}\sigma^2\sub{w}(\hat{A})\theta^2+O(\theta^3)$, $\mathrm{Im}\sigma^2\sub{w}(\hat{A})$ also appears in the width change of the wave packet in the $\hat{\varXi}$ basis. 
These relations are represented on the quadrature phase plane in Fig.~\ref{fig:theory2}(b).
When $\mathrm{Im}\sigma\sub{w}^2(\hat{A})>0$, the wave packet spreads along the $\varOmega$ (45$\degree$) axis, while it narrows along the $\varXi$ (135$\degree$) axis.
This relationship also satisfies the Kennard--Robertson uncertainty relation up to the quadratic of $\theta$: $\sigma\sub{f}^2(\hat{\varOmega})\sigma\sub{f}^2(\hat{\varXi})=1/4+O(\theta^3)$.

\section{Experimental demonstration of weak variances}

\begin{figure}
\begin{center}
\includegraphics[width=8.5cm]{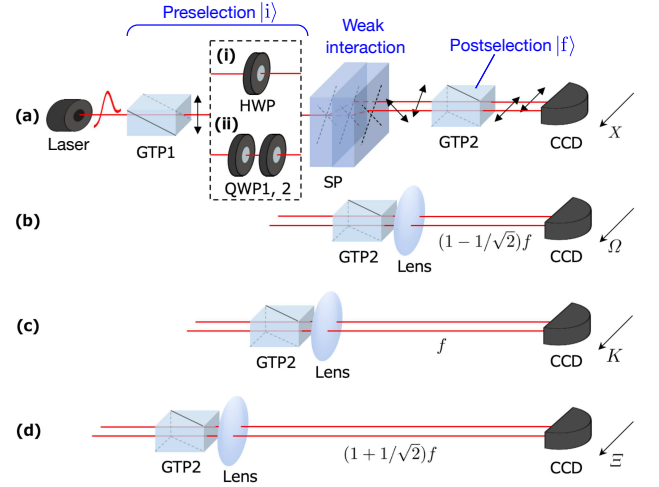} 
\end{center} 
\caption{Experimental setup for observing weak variances.
GTP: Glan--Thompson prism;
HWP: half-wave plate;
QWP: quarter-wave plate;
SP: Savart plate;
CCD: charge-coupled device.
(a) Experimental setup for measuring the probe system in the $\hat{X}$ basis.
 In the pre-selection, the HWP and two QWPs are used to prepare the weak variance to be (i) negative real and (ii) positive purely imaginary, respectively.
(b), (c), (d) Experimental setups for measuring the probe system in the $\hat{\varOmega}$, $\hat{K}$, and $\hat{\varXi}$ bases, respectively.
The lens (focal length $f=1\,\mathrm{m}$) and free-space propagation perform a fractional Fourier transform on the transverse distribution of the beam. 
}\label{fig:4}
\end{figure}

To verify the effects of the weak variance, we experimentally observed the weak variance in the optical system shown in Fig.~\ref{fig:4}. 
In this setup, the target and probe systems were the polarization and transverse spatial modes, respectively, of the laser beam with a central wavelength of 780\,nm (Menlo Systems C-fiber 780).
The polarization mode was a two-state system spanned by (for example) the horizontal--vertical polarization basis $\{\ket{\HH},\ket{\VV}\}$ or the diagonal ($45^\circ$)--antidiagonal ($135^\circ$) polarization basis $\{\ket{\DD}:=(\ket{\HH}+\ket{\VV})/\sqrt{2},\ket{\AA}:=(\ket{\HH}-\ket{\VV})/\sqrt{2}\}$. 
The pre- and post-selection $\{\ket{\ii},\ket{\ff}\}$ in the polarization mode was prepared using Glan--Thompson prisms (GTPs), a half-wave plate (HWP), and quarter-wave plates (QWPs). 
The initial transverse distribution of the beam's amplitude was prepared as a Gaussian distribution $\phi(X)=\pi^{-1/4}\exp(-X^2/2)$, where $X$ is the dimensionless position variable normalized by the standard deviation of this distribution. 
The weak interaction $\exp(-\ii\theta\hat{A}\otimes\hat{K})$ was implemented using a Savart plate (SP), which comprises two orthogonal birefringent crystals ($\upbeta$-BaB$_2$O$_4$, 1-\,mm thickness).
In our setup, $\hat{A}$ was chosen as $\hat{A}=\ket{\DD}\bra{\DD}-\ket{\AA}\bra{\AA}$ and the SP transversely shifted the diagonally (antidiagonally) polarized beam by a distance of $\theta$ ($-\theta$). 
The probe system was finally measured in the $\hat{X}$, $\hat{\varOmega}$, $\hat{K}$, and $\hat{\varXi}$ bases.
In the $\hat{X}$ basis, the transverse intensity distribution of the beam was measured using a charge-coupled device (CCD) camera (Teledyne Princeton Instruments ProEM-HS:512BX3), as shown in Fig.~\ref{fig:4}(a).
The intensity measurements in the $\hat{\varOmega}$, $\hat{K}$, and $\hat{\varXi}$ bases were implemented by fractional Fourier transforming (see Appendix~\ref{sec:AppB} for details) the beam distribution using a lens (focal length $f=1\,\mathrm{m}$) before $\hat{X}$ measurement using the CCD camera, as shown in Figs.~\ref{fig:4}(b)--(d).

To independently verify the effects of the real and imaginary parts of the weak variance, we chose the pre- and post-selected polarization states $\{\ket{\ii},\ket{\ff}\}$ giving (i) negative real and (ii) positive purely imaginary weak variances.
The pre-selected state $\ket{\ii}$ in case (i) was prepared by rotating the fast axis of the HWP through angle $\vartheta_\HH$ from the vertical direction and passing the vertically polarized beam through the rotated HWP. 
The output state became $\ket{\ii}=\cos(2\vartheta_\HH-\pi/4)\ket{\DD}+\sin(2\vartheta_\HH-\pi/4)\ket{\AA}$.
The post-selected state was fixed at $\ket{\ff}=\ket{\HH}$. 
The weak value and weak variance respectively became the following real numbers:
\begin{align}
\bracket{\hat{A}}\sub{w}=\frac{\cos(2\vartheta_\HH)}{\sin(2\vartheta_\HH)},\quad
\sigma^2\sub{w}(\hat{A})=
-\frac{\cos(4\vartheta\sub{H})}{\sin^2(2\vartheta\sub{H})}.\label{eq:15}
\end{align}
The pre-selected state $\ket{\ii}$ in case (ii) was prepared by rotating the fast axis of QWP1 through angle $\vartheta_\mathrm{Q}$ from the vertical direction and passing the vertically polarized beam through the rotated QWP1 and QWP2 (whose fast axis was fixed in the vertical direction). 
The output state became $\ket{\ii}=\cos(\vartheta\sub{Q}-\pi/4)\ket{\DD}+\ee^{-\ii 2\vartheta\sub{Q}}\sin(\vartheta\sub{Q}-\pi/4)\ket{\AA}$. 
The post-selected state was fixed at $\ket{\ff}=\ket{\HH}$, as in case (i).
The weak variance became the following purely imaginary number:
\begin{align}
\sigma^2\sub{w}(\hat{A})=2\ii \frac{\cos(2\vartheta\sub{Q})}{\sin^2(2\vartheta\sub{Q})}.\label{eq:16}
\end{align}

\begin{figure}
\begin{center}
\includegraphics[width=8.5cm]{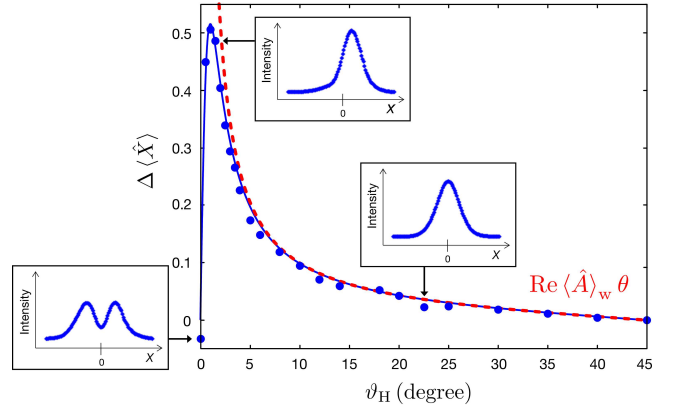} 
\end{center} 
\caption{Measurement results of $\Delta\bracket{\hat{X}}$ in case (i).
The solid blue curve is the theoretical fitting to the measured data (blue dots), and the
red-dashed curve plots the theoretical weak value $\mathrm{Re}\bracket{\hat{A}}\sub{w}\theta$ ($\theta=3.62\times10^{-2}$) versus $\vartheta\sub{H}$.
Insets show the intensity distributions of the beam for several values of $\vartheta\sub{H}$.
When $\vartheta\sub{H}$ is close to zero, the $O(\theta^3)$ term in Eq.~(\ref{eq:1}) dominates and the distribution differs from a Gaussian one.
}\label{fig:5}
\end{figure}

First, we observed the weak value in the transverse displacement of the beam's intensity distribution.
Figure \ref{fig:5} plots the measured displacement of the mean of the beam's intensity distribution in the $\hat{X}$ basis, given by $\Delta\bracket{\hat{X}}:=\bracket{\hat{X}}\sub{f}-\bracket{\hat{X}}\sub{i}$, as a function of $\vartheta\sub{H}$ in case (i). 
When $\vartheta\sub{H}$ is small, the pre- and post-selected states are nearly orthogonal and $\Delta\bracket{\hat{X}}$ becomes large.
The theoretical curve of $\Delta\bracket{\hat{X}}$ (blue solid curve in Fig.~\ref{fig:5}) was fitted to the measured values. 
The fitting parameters were the coupling strength $\theta$, visibility $V$, and the technical error of the rotation angle of HWP $\varDelta\in[-0.5\degree,0.5\degree]$ (for details, see Appendix~\ref{sec:AppE}). 
From the fitting, we determined $\theta=3.62\times10^{-2}$, $V=1.00$, $\varDelta=(1.57\times10^{-3})\degree$.
The weak value $\mathrm{Re}\bracket{\hat{A}}\sub{w}\theta$ ($\theta=3.62\times10^{-2}$) versus $\vartheta\sub{H}$ is also plotted in Fig.~\ref{fig:5} (red-dashed curve).
Most of the measured values were consistent with the theoretical curve, verifying the measurements of the weak values.

\begin{figure}
\begin{center}
\includegraphics[width=8.5cm]{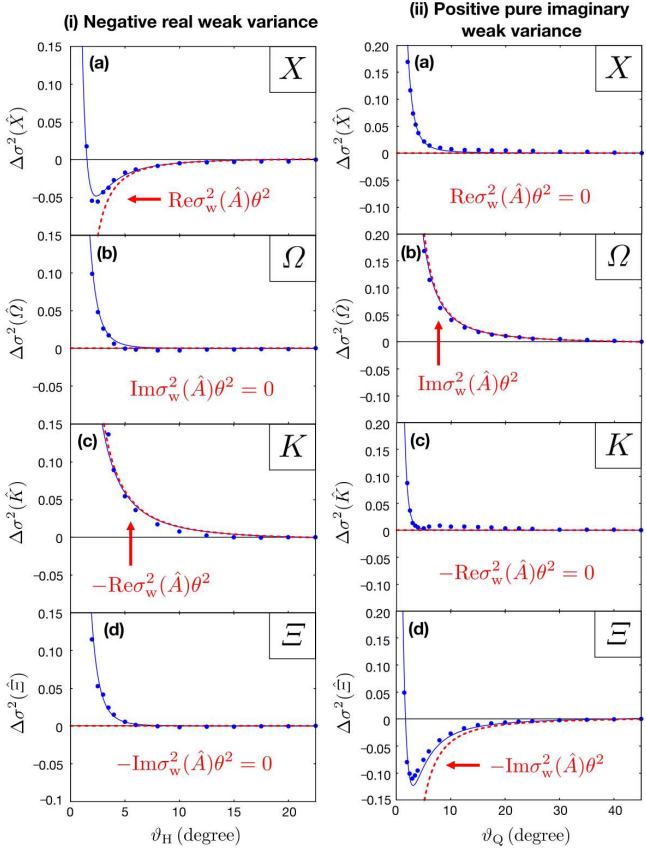} 
\end{center} 
\caption{Measurement results of $\Delta\sigma^2(\hat{M})$ in the cases of (i) negative real weak variance and (ii) purely imaginary weak variance. In panels
(a), (b), (c), and (d), the measurement bases were $\hat{X}$, $\hat{\varOmega}$, $\hat{K}$, and $\hat{\varXi}$, respectively. 
The solid blue lines are the theoretical fittings to the measured data (blue dots), and the
red-dashed curves plot the theoretical weak variance $[\cos(2\alpha)\mathrm{Re}\sigma\sub{w}^2(\hat{A})+\sin(2\alpha)\mathrm{Im}\sigma\sub{w}^2(\hat{A})]\theta^2$ versus $\vartheta\sub{H}$ or $\vartheta\sub{Q}$ (the $\theta$s were obtained by fitting the blue solid lines to the data).
When $\vartheta\sub{H}$ and $\vartheta\sub{Q}$ approach zero, the $O(\theta^3)$ term in Eq.~(\ref{eq:2}) dominates and the measured data notably deviate from the red-dashed lines.
}\label{fig:6}
\end{figure}

We then observed the weak variance, which manifests as the changing variance of the beam's intensity distribution.
Here $\Delta\sigma^2(\hat{M}):=[\sigma^2\sub{f}(\hat{M})-\sigma^2\sub{i}(\hat{M})]/\sigma^2\sub{i}(\hat{M})$ denotes the rate of variance change of the beam's intensity distribution from its initial value in the $\hat{M}=\hat{X}\cos\alpha+\hat{K}\sin\alpha$ basis.
Figure \ref{fig:6} plots the measured $\Delta\sigma^2(\hat{M})$ ($\hat{M}=\hat{X},\hat{\varOmega},\hat{K},\hat{\varXi}$) as functions of $\vartheta\sub{H}$ and $\vartheta\sub{Q}$ in cases (i) and (ii), respectively.
When $\vartheta\sub{H}$ and $\vartheta\sub{Q}$ were small, the pre- and post-selected states were close to orthogonal and the variance changes were large.
The theoretical curves of $\Delta\sigma^2(\hat{M})$ (blue solid curves in Fig.~\ref{fig:6}) were fitted to the measured values using $\theta$, $V$, $\varDelta$, and the intensity of the background light $N$ as fitting parameters (see Appendix~\ref{sec:AppE} for details).
The theoretical weak variances $\left[\cos(2\alpha)\mathrm{Re}\sigma\sub{w}^2(\hat{A})+\sin(2\alpha)\mathrm{Im}\sigma\sub{w}^2(\hat{A})\right]\theta^2$ (red-dashed curves) are also plotted as functions of $\vartheta\sub{H}$ or $\vartheta\sub{Q}$ in Fig.~\ref{fig:6}. 
Again, most of the measured values were consistent with the theoretical curves.
In case (i), $\Delta\sigma^2(\hat{X})$ became negative because the weak variance was a negative real value; correspondingly, the $\Delta\sigma^2(\hat{K})$ increased.
However, $\Delta\sigma^2(\hat{\varOmega})$ and $\Delta\sigma^2(\hat{\varXi})$ remained almost zero because the imaginary part of the weak variance was zero.
In case (ii), where the weak variance was positive and purely imaginary, $\Delta\sigma^2(\hat{\varOmega})$ and $\Delta\sigma^2(\hat{\varXi})$ became positive and negative, respectively.
However, $\Delta\sigma^2(\hat{X})$ and $\Delta\sigma^2(\hat{K})$ remained almost zero.
Thus, the real and imaginary parts of the weak variance appeared as width changes of the wave packet, in accordance with our theory.

\section{Weak variance as a statistic of the weak-valued probability distribution}

\begin{figure}
\includegraphics[width=8.5cm]{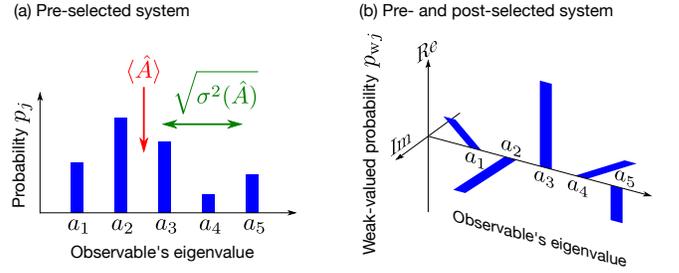}
\caption{(a) Probability distribution of the projective measurement of $\hat{A}$ in a pre-selected system, with an
expectation value and variance of $\bracket{\hat{A}}$ and $\sigma^2(\hat{A})$, respectively.
(b) Weak-valued probability distribution of the observable $\hat{A}$ in pre- and post-selected systems.
The weak-valued probabilities $p_{\mathrm{w}j}$ are complex and expressed on the complex plane.
$\bracket{\hat{A}}\sub{w}$ and $\sigma\sub{w}^2(\hat{A})$ are also complex but cannot be depicted on this graph.
}\label{fig:1}
\end{figure}

In this section, we interpret the weak variance as a statistic of the weak-valued probability distribution, which is a pseudo-probability distribution in the pre- and post-selected system.
This relation is similar to that the variance is expressed as a statistic of the probability distribution in the pre-selected system. 
This statistical interpretation of the weak variance, together with the operational interpretation described above, rationalizes the definition of the weak variance as a counterpart of the variance in pre- and post-selected systems.

We define the weak-valued probability $p_{\mathrm{w}j}:=\bracket{\hat{\varPi}_j}\sub{w}$ as the weak value of each element of the set of projection operators $\{\hat{\varPi}_j\}_j$ that satisfy the completeness condition $\sum_j\hat{\varPi}_j=\hat{1}$.
Weak-valued probabilities can be any complex number outside $[0,1]$, but their sum for all $j$ is unity: $\sum_jp_{\mathrm{w}j}=1$.
By regarding the weak-valued probability as a quantity corresponding to the probability of finding a pre- and post-selected particle in the eigenspace $\hat{\varPi}_j$ between the pre- and post-selection, researchers have found a probabilistic approach to fundamental problems in quantum mechanics \cite{aharonov1991complete,resch2004experimental,PhysRevLett.102.020404,yokota2009direct,denkmayr2014observation,okamoto2016experimental,PhysRevLett.111.240402}. 
Because of their negativity and nonreality, weak-valued probabilities have played an essential role in studies such as the investigation of the relationship between disturbance and complementarity in quantum measurements \cite{wiseman2003directly,mir2007double}, the explanation of the violation of Bell inequality using negative probabilities \cite{PhysRevA.91.012113}, quantum enhancement of the phase estimation sensitivity via post-selection \cite{arvidsson1903quantum}, and understanding out-of-time-order correlators as witnesses of quantum scrambling \cite{PhysRevA.97.042105,PhysRevLett.122.040404}.

The weak value $\bracket{\hat{A}}\sub{w}$ and weak variance $\sigma\sub{w}^2(\hat{A})$ of observable $\hat{A}=\sum_ja_j\hat{\varPi}_j$ can be expressed in terms of the weak-valued probabilities $\{p_{\mathrm{w}j}\}_j$ as follows:
\begin{align}
\bracket{\hat{A}}\sub{w}&=\sum_ja_jp_{\mathrm{w}j}\\
\sigma\sub{w}^2(\hat{A})
&=\sum_j(a_j-\bracket{\hat{A}}\sub{w})^2p_{\mathrm{w}j}
=\sum_j|a_j-\bracket{\hat{A}}\sub{w}|^2p_{\mathrm{w}j},\label{eq:19}
\end{align}
where the second equation in Eq.~(\ref{eq:19}) holds when $\hat{A}$ is Hermitian. 
These expressions are similar to those of the expectation value $\bracket{\hat{A}}=\sum_ja_jp_j$ and variance $\sigma^2(\hat{A})=\sum_j(a_j-\bracket{\hat{A}})^2p_j$, respectively, obtained using probability distribution $\{p_j\}_j$.
In this sense, the weak value and weak variance can be regarded as the expectation value and variance of a weak-valued probability distribution, respectively.
In addition, as the weak-valued probability represents the conditional pseudo-probability of the Kirkwood--Dirac distribution \cite{PhysRev.44.31,RevModPhys.17.195}, the weak value and weak variance are also regarded as the conditional pseudo-expectation value and conditional pseudo-variance of the Kirkwood--Dirac distribution, respectively (see Appendix~\ref{sec:AppF} for details).
Furthermore, the weak value and weak variance satisfy the equations similar to the laws of total expectation and total variance, respectively:
\begin{align}
\bracket{\hat{A}}&=\bracketii{\ii}{\hat{A}}{\ii}=\sum_j|\bracketi{\ff_j}{\ii}|^2\bracket{\hat{A}}_{\mathrm{w}j},\\
\sigma^2(\hat{A})&=\sum_j|\bracketi{\ff_j}{\ii}|^2\sigma_{\mathrm{w}j}^2(\hat{A})+\sum_j|\bracketi{\ff_j}{\ii}|^2\left(\bracket{\hat{A}}_{\mathrm{w}j}-\bracket{\hat{A}}\right)^2,
\end{align}
where $\bracket{\hat{A}}_{\mathrm{w}j}:=\bracketii{\ff_j}{\hat{A}}{\ii}/\bracketi{\ff_j}{\ii}$ and $\sigma_{\mathrm{w}j}^2(\hat{A}):=\bracket{\hat{A}^2}_{\mathrm{w}j}-\bracket{\hat{A}}_{\mathrm{w}j}^2$

Thus far, several definitions of the quantity corresponding to the variance in pre- and post-selected systems have been considered. Examples are the weak variance expressed by Eq.~(\ref{eq:3}) \cite{PhysRevA.52.2538,reznik1995interaction,tanaka2002semiclassical,brodutch2008weak,parks2014weak}, its absolute value \cite{PhysRevA.41.11,parks2018weak}, its real part \cite{feyereisen2015weak,hofer2017quasi}, and other forms \cite{hofmann2011characterization,hofmann2011role,pati2014uncertainty,song2015uncertainty}.
The measurement method (indirect measurement) and statistical expression (a weak-valued probability distribution) of our proposed weak variance are similar to those of the conventional variance.
Therefore, the weak variance defined by Eq.~(\ref{eq:3}) can be regarded as a reasonable counterpart of the variance in pre- and post-selection systems.

\section{Conclusion}

We introduced the weak variance $\sigma\sub{w}^2(\hat{A})$ as a complex counterpart of the variance in pre- and post-selected systems.
We theoretically showed that the weak variance appears as the changing width of the probe wave packet during indirect measurements of pre- and post-selected systems and experimentally demonstrated the weak variance in an optical setup.
We also expressed the weak value $\bracket{\hat{A}}\sub{w}$ and weak variance $\sigma^2\sub{w}(\hat{A})$ as statistics of the weak-valued probability distribution $\{p_{\mathrm{w}j}\}_j$.
These operational and statistical interpretations are similar to the expectation value $\bracket{\hat{A}}$ and variance $\sigma^2(\hat{A})$ in pre-selected systems.
Therefore, our formulation of the weak variance can be considered a reasonable definition of a counterpart of the variance in pre- and post-selected systems.

Extending the concept of the weak variance, we then
defined the \textit{$n$-th order weak moment} of the observable $\hat{A}$ as $\bracket{\hat{A}^n}\sub{w}$.
The set of weak moments $\{\bracket{\hat{A}^k}\sub{w}\}_{k=1}^{n}$ fully characterizes the weak-valued probability distribution $\{p_{\mathrm{w}j}\}_{j=1}^{n}$. A similar relation exists between the set of moments $\{\bracket{\hat{A}^k}\}_{k=1}^{n}$ and the probability distribution $\{p_j\}_{j=1}^n$.
The $n$-th order weak moment $\bracket{\hat{A}^n}\sub{w}$ can be experimentally observed in the indirect measurement setup by including terms up to the $n$-th order of $\theta$ in Eq.~(\ref{eq:5}).
Although the physical meanings of the weak moment is as elusive as the weak value, the weak moment may provide a new perspective on fundamental problems in quantum mechanics.
For example, Scully \textit{et al.}'s claim that the momentum disturbance associated with which-way measurement in Young's double-slit experiment can be avoided \cite{scully1991quantum} has been justified by the negativity of the weak-valued probabilities corresponding to the momentum disturbance, which consequently have zero variance \cite{wiseman2003directly,mir2007double}.
These studies are implicitly based on the weak variance (second-order weak moment) concept.
Similarly, the weak moment is expected to play an important role in other problems of this type.
In addition, measurement methods other than weak measurements with Gaussian probes---such as weak measurements using a qubit probe \cite{wu2009weak} and methods without a probe \cite{PhysRevA.101.042117}---may find new implications for the weak moments.

Finally, as an application of the weak moments $\bracket{\hat{A}^n}\sub{w}$, we propose controlling the probe wave packet by pre- and post-selection of the target system.
In several studies, the probe wave packet was narrowed by appropriate pre- and post-selection of the target system in the weak measurement setup \cite{parks2011variance,de2015uncertainty,matsuoka2017generation}.
If the higher-order weak moments in the $O(\theta^2)$ term of Eq.~(\ref{eq:5}) are properly controlled, we can configure any waveform of the probe (see Appendix~\ref{sec:AppG} for details).
For example, our method may represent a new construction method for the realization of the non-Gaussian states in the quadrature amplitude of light, such as
the cat state \cite{PhysRevLett.57.13,PhysRevA.44.2172} and the Gottesman--Kitaev--Preskill state \cite{PhysRevA.64.012310}, which play important roles in quantum optics.

\begin{acknowledgments}
This research was supported by JSPS KAKENHI Grant Number 16K17524 and 19K14606, the Matsuo Foundation, and the Research Foundation for Opto-Science and Technology.
\end{acknowledgments}

\appendix

\begin{widetext}
\section{Change of probe wave packet in indirect measurements of mixed pre- and post-selected systems}\label{sec:appA}

We calculate the expectation value $\bracket{\hat{M}}\sub{f}$ and variance $\sigma^2\sub{f}(\hat{M})$ of the probe wave packet in indirect measurements, when the pre- and post-selected states are mixed states represented by density operators $\hat{\rho}\sub{i}$ and $\hat{\rho}\sub{f}$, respectively.
$\hat{\rho}\sub{i}=\ket{\ii}\bra{\ii}$ and $\hat{\rho}\sub{f}=\ket{\ff}\bra{\ff}$ corresponds to the case of the pure pre- and post-selected states, and $\hat{\rho}\sub{f}=\hat{1}/d$ the case of the pre-selection alone.
The time evolution of the entire state is calculated as
\begin{align}
&\hat{\rho}\sub{i}\otimes\ket{\phi}\bra{\phi}\no\\
&\xrightarrow{\mathrm{Interaction}}
\exp(-\ii\theta\hat{A}\otimes\hat{K})
\left(\hat{\rho}\sub{i}\otimes\ket{\phi}\bra{\phi}\right)
\exp(\ii\theta\hat{A}\otimes\hat{K})\no\\
&\xrightarrow{\mathrm{Post\text{-}selection}}
\tr\sub{t}\left[
\hat{\rho}\sub{f}
\exp(-\ii\theta\hat{A}\otimes\hat{K})
\left(\hat{\rho}\sub{i}\otimes\ket{\phi}\bra{\phi}\right)
\exp(\ii\theta\hat{A}\otimes\hat{K})
\right]\no\\
&\hspace{1.9cm}=
\tr\left(\hat{\rho}\sub{f}\hat{\rho}\sub{i}\right)
\left[
\ket{\phi}\bra{\phi}
+\left(-\ii\theta\bracket{\hat{A}}\sub{w}\hat{K}\ket{\phi}\bra{\phi}
-\frac{1}{2}\theta^2\bracket{\hat{A}^2}\sub{w}\hat{K}^2\ket{\phi}\bra{\phi}\right)
+\Hc
+\theta^2\tilde{A}\hat{K}\ket{\phi}\bra{\phi}\hat{K}
\right]
+O(\theta^3)\no\\
&\hspace{1.9cm}=:\tilde{\hat{\rho}}_\phi,
\end{align}
where $\tr\sub{t}$ denotes the partial trace in the target system and $\Hc$ represents the Hermitian conjugate of the preceding term.
$\bracket{\hat{A}}\sub{w}={\tr(\hat{\rho}\sub{f}\hat{A}\hat{\rho}\sub{i})}/{\tr(\hat{\rho}\sub{f}\hat{\rho}\sub{i})}$ is the weak value of $\hat{A}$ in the pre- and post-selected systems represented by density operators $\hat{\rho}\sub{i}$ and $\hat{\rho}\sub{f}$, respectively.
We now define $\tilde{A}:={\tr(\hat{\rho}\sub{f}\hat{A}\hat{\rho}\sub{i}\hat{A})}/{\tr\left(\hat{\rho}\sub{f}\hat{\rho}\sub{i}\right)}$. When $\hat{\rho}\sub{i}$ and $\hat{\rho}\sub{f}$ are pure, $\tilde{A}=|\bracket{\hat{A}}\sub{w}|^2$, and when $\hat{\rho}\sub{f}=\hat{1}/d$ (a completely mixed state), $\tilde{A}=\tr(\hat{\rho}\sub{i}\hat{A}^2)=\bracket{\hat{A}^2}$.

The expectation value of $\hat{M}$ for the non-normalized probe state $\tilde{\hat{\rho}}_\phi$ is expressed as $\bracket{\hat{M}}\sub{f}=\tr(\tilde{\hat{\rho}}_\phi\hat{M})/\tr(\tilde{\hat{\rho}}_\phi)$.
The numerator $\tr(\tilde{\hat{\rho}}_\phi\hat{M})$ is calculated as
\begin{align}
&\tr(\tilde{\hat{\rho}}_\phi\hat{M})
=
\tr\left(\hat{\rho}\sub{f}\hat{\rho}\sub{i}\right)
\left[
\bracket{\hat{M}}
+
\left(
-\ii\theta\bracket{\hat{A}}\sub{w}\bracket{\hat{M}\hat{K}}
-\frac{1}{2}\theta^2\bracket{\hat{A}^2}\sub{w}\bracket{\hat{M}\hat{K}^2}
\right)
+\cc
+\theta^2\tilde{A}\bracket{\hat{K}\hat{M}\hat{K}}
\right]
+O(\theta^3),
\end{align}
where $\cc$ represents the complex conjugate of the preceding term.
Because the expectation value of the product of odd numbers of $\hat{X}$ or $\hat{K}$ in our Gaussian probe state becomes zero, for $\hat{M}=\hat{X}\cos\alpha+\hat{K}\sin\alpha$, the above equation can be reduced to
\begin{align}
&\tr(\tilde{\hat{\rho}}_\phi\hat{M})
=
\tr\left(\hat{\rho}\sub{f}\hat{\rho}\sub{i}\right)
\left[
\left(-\ii\theta\bracket{\hat{A}}\sub{w}\bracket{\hat{M}\hat{K}}\right)
+\cc
\right]
+O(\theta^3).
\end{align}
By a similar calculation, the denominator $\tr(\tilde{\hat{\rho}}_\phi)$ is obtained as
\begin{align}
\tr(\tilde{\hat{\rho}}_\phi)
=
\tr\left(\hat{\rho}\sub{f}\hat{\rho}\sub{i}\right)
\left[
1+
\left(-\frac{1}{2}\theta^2\bracket{\hat{A}^2}\sub{w}\bracket{\hat{K}^2}\right)
+\cc
+\theta^2\tilde{A}\bracket{\hat{K}^2}
\right]
+O(\theta^3).
\end{align}
Therefore, the expectation value $\bracket{\hat{M}}\sub{f}$ is expressed as 
\begin{align}
 \bracket{\hat{M}}\sub{f}
=\frac{\tr(\tilde{\hat{\rho}}_\phi\hat{M})}
{\tr(\tilde{\hat{\rho}}_\phi)}
=\left(-\ii\theta\bracket{\hat{A}}\sub{w}\bracket{\hat{M}\hat{K}}\right)
+\cc
+O(\theta^3),
\end{align}
where we have used the following formula:
\begin{align}
\frac{a_0+a_1\theta+a_2\theta^2+O(\theta^3)}
{b_0+b_1\theta+b_2\theta^2+O(\theta^3)}
=
\frac{a_0}{b_0}
+
\frac{a_1b_0-a_0b_1}{b_0^2}\theta
+
\frac{a_2b_0^2-a_0b_0b_2-a_1b_0b_1+a_0b_1^2}{b_0^3}\theta^2
+O(\theta^3).
\end{align}
Because $\bracket{\hat{M}\hat{K}}=(\ii\cos\alpha+\sin\alpha)/2$ in our Gaussian probe state, we obtain a concrete form of $\bracket{\hat{M}}\sub{f}$ as follows:
\begin{align}
 \bracket{\hat{M}}\sub{f}
=\frac{1}{2}\theta\bracket{\hat{A}}\sub{w}(\cos\alpha-\ii\sin\alpha)
+\cc
+O(\theta^3)
=\theta\left(
\mathrm{Re}\bracket{\hat{A}}\sub{w}\cos\alpha
+\mathrm{Im}\bracket{\hat{A}}\sub{w}\sin\alpha
\right)
+O(\theta^3),\label{eq:17}
\end{align}
which matches Eq.~(\ref{eq:1}) in the main text.

The variance of $\hat{M}$ for the non-normalized probe state $\tilde{\hat{\rho}}_\phi$ is expressed as $\sigma^2\sub{f}(\hat{M})=\bracket{\hat{M}^2}\sub{f}-\bracket{\hat{M}}^2\sub{f}$.
The first term is calculated as
\begin{align}
\bracket{\hat{M}^2}\sub{f}
&=\frac{\tr(\tilde{\hat{\rho}}_\phi\hat{M}^2)}
{\tr(\tilde{\hat{\rho}}_\phi)}\no\\
&=\frac{\bracket{\hat{M}^2}
+\left(-\frac{1}{2}\theta^2\bracket{\hat{A}^2}\sub{w}\bracket{\hat{M}^2\hat{K}^2}\right)
+\cc+\theta^2\tilde{A}\bracket{\hat{K}\hat{M}^2\hat{K}}+O(\theta^3)}
{1+
\left(-\frac{1}{2}\theta^2\bracket{\hat{A}^2}\sub{w}\bracket{\hat{K}^2}\right)+\cc+\theta^2\tilde{A}\bracket{\hat{K}^2}+O(\theta^3)}\no\\
&=\bracket{\hat{M}^2}+
\left(-\frac{1}{2}\theta^2\bracket{\hat{A}^2}\sub{w}\bracket{\hat{M}^2\hat{K}^2}\right)
+\cc
+\theta^2\tilde{A}\bracket{\hat{K}\hat{M}^2\hat{K}}
+\bracket{\hat{M}^2}
\left(\frac{1}{2}\theta^2\bracket{\hat{A}^2}\sub{w}\bracket{\hat{K}^2}
+\cc
-\theta^2\tilde{A}\bracket{\hat{K}^2}\right)
+O(\theta^3).
\end{align}
Because the following equations hold for our Gaussian probe state:
\begin{align}
 \bracket{\hat{M}^2}=\bracket{\hat{K}^2}=\frac{1}{2},
\quad
\bracket{\hat{M}^2\hat{K}}=\frac{1}{4}-\frac{1}{2}\left[\cos(2\alpha)-\ii\sin(2\alpha)\right],
\quad
\text{and}\quad
\bracket{\hat{K}\hat{M}^2\hat{K}}=\frac{3}{4},
\end{align}
we obtain the following expression:
\begin{align}
\bracket{\hat{M}^2}\sub{f}
=\frac{\tr(\tilde{\hat{\rho}}_\phi\hat{M}^2)}
{\tr(\tilde{\hat{\rho}}_\phi)}
=\frac{1}{2}
+\frac{1}{2}\theta^2\cos(2\alpha)\mathrm{Re}\bracket{\hat{A}^2}\sub{w}
+\frac{1}{2}\theta^2\sin(2\alpha)\mathrm{Im}\bracket{\hat{A}^2}\sub{w}
+\frac{1}{2}\theta^2\tilde{A}
+O(\theta^3).
\end{align}
The second term $\bracket{\hat{M}}\sub{f}^2$ is calculated using Eq.~(\ref{eq:17}) as
\begin{align}
\bracket{\hat{M}}\sub{f}^2
=\frac{1}{2}\theta^2
\left[\left(\mathrm{Re}\bracket{\hat{A}}\sub{w}\right)^2
-\left(\mathrm{Im}\bracket{\hat{A}}\sub{w}\right)^2\right]
\cos(2\alpha)
+
\theta^2\mathrm{Re}\bracket{\hat{A}}\sub{w}\mathrm{Im}\bracket{\hat{A}}\sub{w}\sin(2\alpha)
+\frac{1}{2}\theta^2|\bracket{\hat{A}}\sub{w}|^2
+O(\theta^3).
\end{align}
Therefore, we obtain the concrete form of $\sigma^2\sub{f}(\hat{M})$ as 
\begin{align}
 \sigma^2\sub{f}(\hat{M})
&=\bracket{\hat{M}^2}\sub{f}-\bracket{\hat{M}}^2\sub{f}\no\\
&=\frac{1}{2}
+\frac{1}{2}\theta^2\cos(2\alpha)\left[
\mathrm{Re}\bracket{\hat{A}^2}\sub{w}-\left(\mathrm{Re}\bracket{\hat{A}}\sub{w}\right)^2+\left(\mathrm{Im}\bracket{\hat{A}}\sub{w}\right)^2
\right]\no\\
&\qquad
+\frac{1}{2}\theta^2\sin(2\alpha)\left[
\mathrm{Im}\bracket{\hat{A}^2}\sub{w}-2\mathrm{Re}\bracket{\hat{A}}\sub{w}\mathrm{Im}\bracket{\hat{A}}\sub{w}
\right]
+\frac{1}{2}\theta^2\left(\tilde{A}-|\bracket{\hat{A}}\sub{w}|^2\right)
+O(\theta^3)\no\\
&=\frac{1}{2}
+\frac{1}{2}\theta^2\cos(2\alpha)\mathrm{Re}\sigma\sub{w}^2(\hat{A})
+\frac{1}{2}\theta^2\sin(2\alpha)\mathrm{Im}\sigma\sub{w}^2(\hat{A})
+\frac{1}{2}\theta^2\left(\tilde{A}-|\bracket{\hat{A}}\sub{w}|^2\right)
+O(\theta^3),\label{eq:18}
\end{align}
where we have used the following formulae:
\begin{align}
\mathrm{Re}\sigma\sub{w}^2(\hat{A})
&=\mathrm{Re}\bracket{\hat{A}^2}\sub{w}-\left(\mathrm{Re}\bracket{\hat{A}}\sub{w}\right)^2+\left(\mathrm{Im}\bracket{\hat{A}}\sub{w}\right)^2,\\
\mathrm{Im}\sigma\sub{w}^2(\hat{A})
&=\mathrm{Im}\bracket{\hat{A}^2}\sub{w}-2\mathrm{Re}\bracket{\hat{A}}\sub{w}\mathrm{Im}\bracket{\hat{A}}\sub{w}.
\end{align}

In particular, when $\hat{\rho}\sub{i}$ and $\hat{\rho}\sub{f}$ are pure, $\tilde{A}=|\bracket{\hat{A}}\sub{w}|^2$, so Eq.~(\ref{eq:18}) matches Eq.~(\ref{eq:2}) in the main text.
However, when $\hat{\rho}\sub{f}=\hat{1}/d$ (pre-selection only), $\tilde{A}=\bracket{\hat{A}^2}$, $\bracket{\hat{A}}\sub{w}=\bracket{\hat{A}}$, and $\sigma\sub{w}^2(\hat{A})=\sigma^2(\hat{A})\in\mathbb{R}$; therefore, we have
\begin{align}
 \sigma^2\sub{f}(\hat{M})
=\frac{1}{2}
+\frac{1}{2}\theta^2\cos(2\alpha)\sigma^2(\hat{A})
+\frac{1}{2}\theta^2\sigma^2(\hat{A})
+O(\theta^3).
\end{align}
When $\alpha=0$, we obtain
\begin{align}
 \sigma^2\sub{f}(\hat{M})
=\sigma^2\sub{f}(\hat{X})
=\frac{1}{2}
+\theta^2\sigma^2(\hat{A})
+O(\theta^3),
\end{align}
which equals $\sigma^2\sub{f}(\hat{X})$ in Eq.~(\ref{eq:4}) of the main text but without the $O(\theta^3)$ term (which vanishes in the full-order expansion in this case).
Note that if the probe state is not a Gaussian wave packet, the expectation value and variance of $\hat{M}$ for the probe wave packet after the post-selection will differ from Eqs.~(\ref{eq:17}) and (\ref{eq:18}), respectively.

\section{Fractional Fourier transform and its optical realization}\label{sec:AppB}

\subsection{Definition of fractional Fourier transform}

For any real number $\alpha$, the $\alpha$-angle fractional Fourier transform of a function $\phi$ is defined as
\begin{align}
 \mathcal{F}_\alpha[\phi](\omega):=
\sqrt{\frac{1-\ii\cot(\alpha)}{2\pi}}
\int_{-\infty}^\infty
\dd x \phi(x)
\exp\left[\ii\left(
\frac{
\cot(\alpha)\omega^2
-2\csc(\alpha)\omega x
+\cot(\alpha)x^2}
{2}
\right)
\right],
\end{align}
where $x$ and $\omega$ are dimensionless variables, $\cot(\alpha)=1/\tan(\alpha)$, and $\csc(\alpha)=1/\sin(\alpha)$.
After transforming function $\phi_0$ by $\mathcal{F}_\alpha$, we obtain a new function $\phi_\alpha$.
When $\alpha=\pi/2$, $\mathcal{F}_\alpha$ reduces to the standard Fourier transform $\mathcal{F}$
\begin{align}
\mathcal{F}_{\pi/2}[\phi_0](\omega)
=\phi_{\pi/2}(\omega)
=\frac{1}{\sqrt{2\pi}}
\int_{-\infty}^\infty\dd x \phi_0(x)\ee^{-\ii\omega x}
=\mathcal{F}[\phi_0](\omega).
\end{align}
The fractional Fourier transforms with $\alpha=\pm\pi/4$ and $\alpha=\pm3\pi/4$ are respectively expressed as
\begin{align}
 \mathcal{F}_{\pm\pi/4}[\phi_0](\omega)
&=\phi_{\pm\pi/4}(\omega)
%
=
\sqrt{\frac{1\mp\ii}{2\pi}}
\int_{-\infty}^\infty\dd x \phi_0(x)
\exp\left[\pm\ii\left(
\frac{
\omega^2
-2\sqrt{2}\omega x
+x^2}
{2}
\right)
\right],\label{eq:12}\\
\mathcal{F}_{\pm 3\pi/4}[\phi_0](\omega)
&=\phi_{\pm3\pi/4}(\omega)
%
=
\sqrt{\frac{1\pm\ii}{2\pi}}
\int_{-\infty}^\infty\dd x \phi_0(x)
\exp\left[\mp\ii\left(
\frac{
\omega^2
+2\sqrt{2}\omega x
+x^2}
{2}
\right)
\right].\label{eq:13}
\end{align}
We call $\mathcal{F}_{\pm\pi/4}$ and $\mathcal{F}_{\pm3\pi/4}$ the $\pm1/2$- and $\pm3/2$-Fourier transform, respectively.


\subsection{Relationship between observables $\hat{X}$, $\hat{K}$, $\hat{\varOmega}$, and $\hat{\varXi}$}

The canonical conjugate of an observable $\hat{X}$, denoted by $\hat{K}$, satisfies the canonical commutation relation $[\hat{X},\hat{K}]=\ii\hat{1}$. 
$\hat{X}$ and $\hat{K}$ are spectrally decomposed as follows:
\begin{align}
 \hat{X}=\int_{-\infty}^\infty\dd X X\ket{X}\bra{X},\quad
\hat{K}=\int_{-\infty}^\infty\dd K K\ket{K}\bra{K}.\label{eq:6}
\end{align}
Their eigenvectors $\ket{X}$ and $\ket{K}$ are interrelated through the Fourier transform:
\begin{align}
 \ket{K}=\mathcal{F}_{-\pi/2}[\ket{X}](K)
=\frac{1}{\sqrt{2\pi}}\int_{-\infty}^\infty\dd X \ket{X}\ee^{\ii KX}.\label{eq:7}
\end{align}
Observables $\hat{\varOmega}$ and $\hat{\varXi}$ are defined as
\begin{align}
 \hat{\varOmega}:=\frac{\hat{X}+\hat{K}}{\sqrt{2}},
\quad
\hat{\varXi}:=\frac{-\hat{X}+\hat{K}}{\sqrt{2}}.\label{eq:9}
\end{align}
They satisfy the canonical commutation relation $[\hat{\varOmega},\hat{\varXi}]=\ii\hat{1}$.
Observables $\hat{\varOmega}$ and $\hat{\varXi}$ are spectrally decomposed as follows:
\begin{align}
 \hat{\varOmega}=\int_{-\infty}^\infty\dd\varOmega\varOmega\ket{\varOmega}\bra{\varOmega},
\quad
\hat{\varXi}=\int_{-\infty}^\infty\dd\varXi\varXi\ket{\varXi}\bra{\varXi}.\label{eq:8}
\end{align}
The eigenvectors $\ket{\varOmega}$ and $\ket{\varXi}$ of $\hat{\varOmega}$ and $\hat{\varXi}$, respectively, are related to $\ket{X}$ by the $-1/2$- and $-3/2$-Fourier transforms, respectively:
\begin{align}
\ket{\varOmega}&=\mathcal{F}_{-\pi/4}[\ket{X}](\varOmega)
=\sqrt{\frac{1+\ii}{2\pi}}\int_{-\infty}^\infty\dd X \ket{X}
\exp\left[-\ii\left(\frac{\varOmega^2-2\sqrt{2}\varOmega X+X^2}{2}\right)\right],\\
\ket{\varXi}&=\mathcal{F}_{-3\pi/4}[\ket{X}](\varXi)
=\sqrt{\frac{1-\ii}{2\pi}}\int_{-\infty}^\infty\dd X \ket{X}
\exp\left[\ii\left(\frac{\varXi^2+2\sqrt{2}\varXi X+X^2}{2}\right)\right].
\end{align}
When state $\ket{\phi}$ is expanded in each basis as $\ket{\phi}=\int_{-\infty}^\infty\dd X\phi_0(X)\ket{X}=\int_{-\infty}^\infty\dd \varOmega\phi_{\pi/4}(\varOmega)\ket{\varOmega}=\int_{-\infty}^\infty\dd K\phi_{\pi/2}(K)\ket{K}=\int_{-\infty}^\infty\dd \varXi\phi_{3\pi/4}(\varXi)\ket{\varXi}$, the relation between each wave function and basis vector is summarized as
\begin{gather}
\phi_0(X)\xrightarrow{\mathcal{F}_{\pi/4}} \phi_{\pi/4}(\varOmega)
\xrightarrow{\mathcal{F}_{\pi/4}} \phi_{\pi/2}(K)
\xrightarrow{\mathcal{F}_{\pi/4}} \phi_{3\pi/4}(\varXi),\\
\ket{X}\xrightarrow{\mathcal{F}_{-\pi/4}} \ket{\varOmega}
\xrightarrow{\mathcal{F}_{-\pi/4}} \ket{K}
\xrightarrow{\mathcal{F}_{-\pi/4}} \ket{\varXi}.
\end{gather}

\subsection{Optical realization of measurement of observables $\hat{X}$, $\hat{K}$, $\hat{\varOmega}$, and $\hat{\varXi}$}

This Appendix describes the optical system for measuring the observables $\hat{X}$, $\hat{K}$, $\hat{\varOmega}$, and $\hat{\varXi}$ for a photon beam with a transverse distribution state $\ket{\phi}$.
To measure the dimensionless transverse-position observable $\hat{X}$, we measure the photon's transverse position using a photon detector with suitable spatial resolution.
To measure $\hat{K}$, we optically Fourier-transform the photon's wavefunction $\phi_0(X)=\bracketi{X}{\phi}$ and measure the transverse position of the resulting function $\phi_{\pi/2}(K)$.
The optical Fourier transform is realized by combining a lens passage with free-space propagation. 
Similarly, $\hat{\varOmega}$ and $\hat{\varXi}$ can be measured by optically $1/2$- and $3/2$-Fourier-transforming $\phi_0(X)$ and measuring the transverse positions of the resulting functions $\phi_{\pi/4}(\varOmega)$ and $\phi_{3\pi/4}(\varXi)$, respectively. 
In what follows, we derive the optical $1/2$- and $3/2$-Fourier transform by combining the lens passage and free-space propagation.

We assume that the beam is propagating in the $z$ direction and define
$x$, $k$, and $k_x$ as the transverse position, total wavenumber, and $x$ component of the wavevector, respectively. 
We apply the paraxial approximation and assume that $k$ does not depend on $k_x$ because $k_x\ll k$.
We then define the dimensionless variables $X:=xk$ and $K_x:=k_x/k$.
Free-space propagation through distance $d$ is represented in wavenumber space by the following transfer function:
\begin{align}
H\sur{free}_D(K_x)
&\propto \exp\left(-\ii\frac{DK_x^2}{2}\right),
\end{align}
where $D:=dk$ is the dimensionless distance. 
In position space, free-space propagation is represented by a convolution with the following function:
\begin{align}
h\sur{free}_D(X)&\propto\int_{-\infty}^\infty\dd K_x H\sur{free}_D(K_x)\ee^{\ii K_xX}
\propto\exp\left(\ii \frac{X^2}{2D}\right).
\end{align}
Meanwhile, passage through a lens with focal length $f$ is represented in the position space by the following transfer function:
\begin{align}
h\sur{lens}_F(X)\propto \exp\left(-\ii \frac{X^2}{2F}\right),
\end{align}
where $F:=fk$ is the dimensionless focal length. 
In wavevector space, passage through this lens is represented by a convolution with the following function:
\begin{align}
H\sur{lens}_F(K_x)
&\propto\int_{-\infty}^\infty\dd X h\sur{lens}_F(X)\ee^{-\ii K_xX}
\propto\exp\left(\ii \frac{FK_x^2}{2}\right).
\end{align}
If a photon with a transverse wave function $\phi_0(X)$ sequentially passes through a lens with focal length $f$, propagates in free space through distance $d$, and passes through another lens with focal length $f$, the resultant wave function is calculated as 
\begin{align}
 \phi_0(X)
&\xrightarrow{\text{lens }f}
\phi_0(X)\exp\left(-\ii\frac{X^2}{2F}\right)\\
&\xrightarrow{\text{free-space propagation }d}
\int_{-\infty}^\infty\dd X'
\phi_0(X')\exp\left(-\ii\frac{X'^2}{2F}\right)
\exp\left[\ii \frac{(X-X')^2}{2D}\right]\\
&\xrightarrow{\text{lens }f}
\int_{-\infty}^\infty\dd X'
\phi_0(X')\exp\left(-\ii\frac{X'^2}{2F}\right)
\exp\left[\ii \frac{(X-X')^2}{2D}\right]
\exp\left(-\ii\frac{X^2}{2F}\right)\no\\
&\hspace{1cm}
=\int_{-\infty}^\infty\dd X'
\phi_0(X')
\exp\left[
\ii\frac{(F-D)X^2-2FXX'+(F-D)X'^2}{2FD}
\right].\label{eq:14}
\end{align}

If we choose $D=F$, we obtain the standard Fourier transform of $\phi_0$: 
\begin{align}
\text{Eq.}~(\ref{eq:14})
&=\int_{-\infty}^\infty\dd X'\phi_0(X')\exp\left(
\ii\frac{-XX'}{F}
\right)
\propto
\mathcal{F}_{\pi/2}[\phi_0]\left(\frac{X}{F}\right)
=\phi_{\pi/2}\left(\frac{X}{F}\right).
\end{align}
After the Fourier transform, the scale of the wave function can be adjusted by adjusting the focal length $F$.
If we choose $D=(1-1/\sqrt{2})F$, then
\begin{align}
\text{Eq.}~(\ref{eq:14})
&=\int_{-\infty}^\infty\dd X'\phi_0(X')\exp\left[
\ii\frac{X^2-2\sqrt{2}XX'+X'^2}{(2-\sqrt{2})F}
\right]\no\\
&=\int_{-\infty}^\infty\dd X'\phi_0(X')
\exp\left\{
\frac{\ii}{2}\left[
\left(\frac{X}{\sqrt{(\sqrt{2}-1)F}}\right)^2
-2\sqrt{2}
\left(\frac{X}{\sqrt{(\sqrt{2}-1)F}}\right)
\left(\frac{X'}{\sqrt{(\sqrt{2}-1)F}}\right)
+\left(\frac{X'}{\sqrt{(\sqrt{2}-1)F}}\right)^2
\right]
\right\}\no\\
&\propto
\mathcal{F}_{\pi/4}\left[\phi_{0,F}^-
\right]
\left(\frac{X}{\sqrt{(\sqrt{2}-1)F}}\right)
\quad\left[\phi_{0,F}^-(X):=\phi_0
\left(X\sqrt{(\sqrt{2}-1)F}\right)
\right]\no\\
&=\phi_{\pi/4,F}^-\left(\frac{X}{\sqrt{(\sqrt{2}-1)F}}\right),
\end{align}
where $\phi_{0,F}^-(X)$ is a scaled wave function of $\phi_0(X)$.
In this manner, we obtain the $1/2$-Fourier transform of $\phi_{F,0}(X)$. 
Similarly, if we choose $D=(1+1/\sqrt{2})F$, the $3/2$-Fourier transform is obtained as follows:
\begin{align}
\text{Eq.}~(\ref{eq:14})
&=\int_{-\infty}^\infty\dd X'\phi_0(X')\exp\left[
-\ii\frac{X^2+2\sqrt{2}XX'+X'^2}{(2+\sqrt{2})F}
\right]\no\\
&=\int_{-\infty}^\infty\dd X'\phi_0(X')
\exp\left\{
-\frac{\ii}{2}\left[
\left(\frac{X}{\sqrt{(\sqrt{2}+1)F}}\right)^2
+2\sqrt{2}
\left(\frac{X}{\sqrt{(\sqrt{2}+1)F}}\right)
\left(\frac{X'}{\sqrt{(\sqrt{2}+1)F}}\right)
+\left(\frac{X'}{\sqrt{(\sqrt{2}+1)F}}\right)^2
\right]
\right\}\no\\
&\propto
\mathcal{F}_{3\pi/4}\left[\phi_{0,F}^+
\right]
\left(\frac{X}{\sqrt{(\sqrt{2}+1)F}}\right)
\quad\left[\phi_{0,F}^+(X):=\phi_0
\left(X\sqrt{(\sqrt{2}+1)F}\right)
\right]\no\\
&=\phi_{3\pi/4,F}^-\left(\frac{X}{\sqrt{(\sqrt{2}+1)F}}\right), 
\end{align}
where $\phi_{0,F}^+(X)$ is a scaled wave function of $\phi_0(X)$.

Note that the second lens, which causes phase modulation in the position space, does not affect the measured intensity (projection) in the position basis.
Therefore, in the experiment (see main text), the intensities of the beam's transverse distribution in the $\hat{X}$, $\hat{\varOmega}$, $\hat{K}$ and $\hat{\varXi}$ bases were measured by inserting only one lens followed by free-space propagation.

\section{Theoretical derivations of the expectation value and variance in our experimental probe system }\label{sec:AppE}

First, we derive the exact formulae of the weak value and weak variance in our experimental setup.
In the experiment, we assumed the pre-selected state $\ket{\ii}$, post-selected state $\ket{\ff}$, and the observable $\hat{A}$ as
\begin{align}
\ket{\ii}&=\cos\frac{\vartheta\sub{i}}{2}\ket{0}+\ee^{\ii\varphi\sub{i}}\sin\frac{\vartheta\sub{i}}{2}\ket{1},\quad
\ket{\ff}=\frac{1}{\sqrt{2}}(\ket{0}+\ket{1}),\quad
\hat{A}=\ket{0}\bra{0}-\ket{1}\bra{1}.
\end{align}
The weak value and weak variance are calculated as
\begin{align}
\bracket{\hat{A}}\sub{w}
&=\frac{\cos\vartheta\sub{i}+\ii\sin\vartheta\sub{i}\sin\varphi\sub{i}}
{1+\sin\vartheta\sub{i}\cos\varphi\sub{i}},\label{eq:20}\\
\sigma^2\sub{w}(\hat{A})
&=
\frac{2\sin\vartheta\sub{i}(\sin\vartheta\sub{i}+\cos\varphi\sub{i})
-\ii\sin(2\vartheta\sub{i})\sin\varphi\sub{i}}
{(1+\sin\vartheta\sub{i}\cos\varphi\sub{i})^2}. \label{wvar}
\end{align}
In our experiment, we used the following values:
\begin{align}
\vartheta\sub{i}=\begin{cases}4\vartheta\sub{H}-\pi/2\quad&[\mathrm{case\ (i)}]\\2\vartheta\sub{Q}-\pi/2\quad&[\mathrm{case\ (ii)}]\end{cases},\quad
\varphi\sub{i}=\begin{cases}0\quad&[\mathrm{case\ (i)}]\\-2\vartheta\sub{Q}\quad&[\mathrm{case\ (ii)}]\end{cases}.\label{eq:21}
\end{align}
Substituting these terms into Eqs.~(\ref{eq:20}) and (\ref{wvar}), we obtain Eqs.~(\ref{eq:15}) and (\ref{eq:16}), respectively.

Next, we derive the theoretical curves of the expectation value and variance of the probe wave packet demonstrated in our experiment.
The state of the entire system after the interaction is
\begin{align}
\exp(-\ii\theta\hat{A}\otimes\hat{K})\ket{\ii}\ket{\phi}
&=\cos\frac{\vartheta\sub{i}}{2}\ket{0}\exp(-\ii\theta \hat{K})\ket{\phi}
+\ee^{\ii\varphi\sub{i}}\sin\frac{\vartheta\sub{i}}{2}\ket{1}\exp(\ii\theta \hat{K})\ket{\phi}.	\label{interact}
\end{align}
For notational simplicity, we denote the first and second terms on the right-hand side of Eq.~(\ref{interact}) by $\ket{\varPhi_0}$ and $\ket{\varPhi_1}$, respectively.
Considering the decrease in visibility $V\in[0,1]$, the state of the entire system after the interaction is expressed by the following density operator: 
\begin{align}
\hat{\rho}:=\ket{\varPhi_0}\bra{\varPhi_0}+\ket{\varPhi_1}\bra{\varPhi_1}+V\left(
\ket{\varPhi_0}\bra{\varPhi_1}+\ket{\varPhi_1}\bra{\varPhi_0}
\right).
\end{align}
After post-selecting the target system onto $\ket{\ff}$, the non-normalized probe state becomes $\tilde{\hat{\rho}}\sub{f}:=\bra{\ff}\hat{\rho}\ket{\ff}$.
The initial probe state is assumed as a Gaussian distribution $\bracketi{X}{\phi}=\phi(X)=\pi^{-1/4}\exp(-X^2/2)$.
The expectation value of the observable $\hat{M}=\hat{X}\cos\alpha+\hat{K}\sin\alpha$ for $\tilde{\hat{\rho}}\sub{f}$ is calculated as 
\begin{align}
\bracket{\hat{M}}\sub{f}
&=\frac{\tr(\tilde{\hat{\rho}}\sub{f}\hat{M})}{\tr(\tilde{\hat{\rho}}\sub{f})}
=\theta
\frac{\cos\alpha\cos\vartheta\sub{i}
+V\sin\alpha\sin\vartheta\sub{i}\sin\varphi\sub{i}\ee^{-\theta^2}}
{1+V\sin\vartheta\sub{i}\cos\varphi\sub{i}\ee^{-\theta^2}}
.\label{eq:10}
\end{align}
To obtain the theoretical curve of the expectation value of $\hat{X}$ in case (i), we substitute Eq.~(\ref{eq:21}) and $\alpha=0$ into Eq.~(\ref{eq:10}) as
\begin{align}
\bracket{\hat{X}}\sub{f}
=
\theta
\frac{\sin(4\vartheta\sub{H})}
{1-V\cos(4\vartheta\sub{H})\ee^{-\theta^2}}.
\end{align}
Here we assume a technical error in the rotation angle of the HWP $\vartheta\sub{H}$ $\varDelta\in[-0.5\degree,0.5\degree]$ that occurs in the experiment; accordingly, the fitting function is given as
\begin{align}
\bracket{\hat{X}}\sub{f}
=
\theta
\frac{\sin[4(\vartheta\sub{H}+\varDelta)]}
{1-V\cos[4(\vartheta\sub{H}+\varDelta)]\ee^{-\theta^2}}.
\end{align}
Fitting this function to the measured data with $\theta$, $V$ and $\varDelta$ as the fitting parameters, we obtained $\theta=3.62\times10^{-2}$, $V=1.00$, and $\varDelta=1.57\times10^{-3}$.

Meanwhile, the theoretical variance curve of the observable $\hat{M}$ for $\tilde{\hat{\rho}}\sub{f}$, $\sigma\sub{f}^2(\hat{M})$, is calculated as follows.
The non-normalized probability density distribution of the signal $\tilde{\hat{\rho}}$ in the measurement basis $\hat{M}=\int_{-\infty}^{\infty}\dd M M \ket{M}\bra{M}$ is given by $I\sub{signal}(M)=\bracketii{M}{\tilde{\hat{\rho}}}{M}$.
We now consider the effect of background light on the measured variances. The background light is modeled as the following rectangular function with intensity $N$ and width $2L$:  
\begin{align}
I\sub{background}(M)=
\begin{cases}
N & (M\in[-L,L])\\
0 & (\mathrm{otherwise}),\\
\end{cases}
\end{align}
where $L=5.6$ in our experimental setup.
The normalized probability density distribution of the summed background and signal intensities is given by
\begin{align}
I\sub{total}(M)
=\frac{I\sub{signal}(M)+I\sub{background}(M)}
{\int_{-\infty}^\infty\dd M\left[I\sub{signal}(M)+I\sub{background}(M)\right]}.
\end{align}
Using $I\sub{total}(M)$, the theoretical curve of $\sigma\sub{f}^2(\hat{M})$ is calculated as follows:
\begin{align}
 \sigma\sub{f}^2(\hat{M})
&=\int_{-\infty}^\infty M^2 I\sub{total}(M)
-\left[\int_{-\infty}^\infty M I\sub{total}(M)\right]^2\no\\
&=
\frac{1}{2}
+\theta^2
\frac{\sin\vartheta\sub{i}
\left\{
\left(\cos^2\alpha-V^2\sin^2\alpha\ee^{-2\theta^2}\right)
\sin\vartheta\sub{i}
+V\ee^{-\theta^2}\left[
\cos(2\alpha)
\cos\varphi\sub{i}
+\sin(2\alpha)\cos\vartheta\sub{i}
\sin\varphi\sub{i}
\right]
\right\}
}
{\left(1+V\sin\vartheta\sub{i}\cos\varphi\sub{i}
\ee^{-\theta^2}+4LN\right)^2}\no\\
&\quad
+4LN\theta^2
\frac{\cos^2\alpha
-V\ee^{-\theta^2}\sin^2\alpha\sin\vartheta\sub{i}\cos\varphi\sub{i}
}
{\left(1+V\sin\vartheta\sub{i}\cos\varphi\sub{i}
\ee^{-\theta^2}+4LN\right)^2}
+\frac{2}{3}LN
\frac{
2L^2-3
}
{1+V\sin\vartheta\sub{i}\cos\varphi\sub{i}
\ee^{-\theta^2}+4LN}.\label{eq:11}
\end{align}
The fitting functions in cases (i) and (ii) are derived by substituting Eq.~(\ref{eq:21}) into Eq.~(\ref{eq:11}) and replacing $\vartheta\sub{H}$ and $\vartheta\sub{Q}$ with $\vartheta\sub{H}+\varDelta$ and $\vartheta\sub{Q}+\varDelta$ ($\varDelta\in[-0.5\degree,0.5\degree]$), respectively.
These functions were fitted to the measured data with $\theta$, $V$, $\varDelta$, and $N$ as the fitting parameters. The fitting results are summarized in Tab.~\ref{tab:1}. 
When fitting the $\hat{X}$ measurement in case (i), $N$ was the only fitting parameter. The other parameters were fixed at $\theta=3.62\times10^{-2}$, $V=1.00$, and $\varDelta=(1.57\times10^{-3})\degree$ because the experimental settings were unchanged from those of the $\bracket{\hat{X}}\sub{f}$ measurements.

\begin{table*}
\caption{Results of fitting $\theta$, $V$, $\varDelta$, and $N$ to $\sigma\sub{f}^2(\hat{M})$
(``--'' denotes a fixed parameter)
}\label{tab:1}
\begin{ruledtabular}
\begin{tabular}{cp{0.1pt}ccccp{0.1pt}cccc}
 &&\multicolumn{4}{c}{Case (i)}&&\multicolumn{4}{c}{Case (ii)}\\
 Measurement basis && $\theta$&$V$&$\varDelta$&$N$&&$\theta$&$V$&$\varDelta$&$N$\\ \hline
 $\hat{X}$&& -- & -- & -- & $2.11\times10^{-6}$ && $5.56\times10^{-2}$ & 1.000 & 0.482$\degree$ & $8.21\times10^{-7}$\\
 $\hat{\varOmega}$ && $5.96\times10^{-2}$ & 1.00 & 0.185$\degree$ & 0.00 && $5.41\times10^{-2}$ & 1.000 & -0.161$\degree$ & 0.00\\
 $\hat{K}$ && $4.92\times10^{-2}$ & 1.00 & 0.500$\degree$ & 0.00 && $2.34\times10^{-2}$ & 0.999 & -0.456$\degree$ & 0.00\\
 $\hat{\varXi}$ && $5.24\times10^{-2}$ & 0.999 & -0.108$\degree$ & 0.00 && $5.39\times10^{-2}$ & 0.999 & 0.407$\degree$ & $2.49\times10^{-6}$\\
\end{tabular}
\end{ruledtabular}
\end{table*}


\section{Weak variance as a conditional pseudo-variance of the Kirkwood--Dirac distribution}\label{sec:AppF}

We show that the weak values and weak variances can be interpreted as conditional pseudo-expectation values and conditional pseudo-variances of the Kirkwood--Dirac (KD) distribution \cite{PhysRev.44.31,RevModPhys.17.195}, respectively.
The $(j,k)$ component of the KD distribution of state $\ket{\ii}$ can be expanded in two orthonormal bases $\{\ket{a_j}\}_j$ and $\{\ket{a'_k}\}_k$ as 
\begin{align}
D(a_j,a'_k|\ii)&:=\tr(\ket{a_j}\bracketi{a_j}{a'_k}\bracketi{a'_k}{\ii}\bra{\ii}).
\end{align}
The KD distribution is a joint pseudo-probability distribution representing the quantum state $\ket{\ii}$ and is generally a complex number.
The KD distributions of states with indices $j, k$ sum to unity: $\sum_{jk}D(a_j,a'_k|\ii)=1$.
The marginal distribution of the KD distribution summed over one index becomes the projection probability distribution of $\ket{\ii}$ in the other basis:
\begin{align}
\sum_jD(a_j,a'_k|\ii)=|\bracketi{a'_k}{\ii}|^2,\quad
\sum_kD(a_j,a'_k|\ii)=|\bracketi{a_j}{\ii}|^2.
\end{align}
When we choose $\ket{a_j}=\ket{\ff}$, the conditional pseudo-probability $D(a'_k|\ii,\ff)$ of the KD distribution becomes
\begin{align}
D(a'_k|\ii,\ff)&:=\frac{D(\ff,a'_k|\ii)}{\sum_k D(\ff,a'_k|\ii)}=\frac{\bracketi{\ff}{a'_k}\bracketi{a'_k}{\ii}}{\bracketi{\ff}{\ii}}=p'_{\mathrm{w}k},
\end{align} 
where $\{p'_{\mathrm{w}k}\}_k$ is the weak-valued probability distribution of the pre- and post-selection system $\{\ket{\ii},\ket{\ff}\}$ in the orthonormal basis $\{\ket{a'_k}\}_k$.
Therefore, the weak value $\bracket{\hat{A}'}\sub{w}$ and weak variance $\sigma^2\sub{w}(\hat{A}')$ of observable $\hat{A}':=\sum_ka'_k\ket{a'_k}\bra{a'_k}$ are represented as the conditional pseudo-expectation values and conditional pseudo-variance of the KD distribution, respectively, as follows: 
\begin{align}
\sum_k a'_k D(a'_k|\ii,\ff)
&=\sum_k a'_k p'_{\mathrm{w}k}
=\bracket{\hat{A}'}\sub{w},\\
\sum_k (a'_k-\bracket{\hat{A}'}\sub{w})^2 D(a'_k|\ii,\ff)
&=\sum_k (a'_k-\bracket{\hat{A}'}\sub{w})^2 p'_{\mathrm{w}k}
=\sigma^2\sub{w}(\hat{A}').
\end{align}


\section{Control of the probe wavefunction by pre- and post-selecting the target system}\label{sec:AppG}

In indirect measurements of pre- and post-selected systems, the probe state after the post-selection can be controlled by appropriately choosing the pre- and post-selected target system.
The wavefunction $\tilde{\phi}(K)$ in the $\hat{K}$ basis of the probe state after the post-selection $\ket{\tilde{\phi}}$ [Eq.~(\ref{eq:5}) in the main text] is expressed for all orders of $\theta$ as
\begin{align}
\tilde{\phi}(K)
=
\bracketi{K}{\tilde{\phi}}
=
\bra{K}
\left[
\bracketi{\ff}{\ii}
\sum_{n=0}^\infty
\frac{(-\ii\theta)^n}{n!}\bracket{\hat{A}^n}\sub{w}\hat{K}^n
\ket{\phi}
\right]
=\bracketi{\ff}{\ii}
\sum_{n=0}^\infty
\frac{(-\ii\theta)^n}{n!}\bracket{\hat{A}^n}\sub{w}K^n
\phi(K).
\end{align}
Let $\phi_\star(K)$ be the wavefunction in the $\hat{K}$ basis of the desired probe state.
To realize $\phi_\star(K)$ (except for a constant multiple), we can choose the weak moments $\{\bracket{\hat{A}^n}\}_n$ so that
\begin{align}
 \sum_{n=0}^\infty
\frac{(-\ii\theta)^n}{n!}\bracket{\hat{A}^n}\sub{w}K^n
\propto
\frac{\phi_\star(K)}{\phi(K)}.
\end{align}
When the target system is $d$-dimensional and $\hat{A}$ has full rank, we can independently choose the values of $d$ weak moments $\bracket{\hat{A}^n}$.
Therefore, by appropriately fixing the weak-moment values of the low-order terms of $\theta$ ($n=1,\cdots,d$), which considerably affect the waveform, we can approximate the desired wavefunction.

\end{widetext}

\end{spacing}

\nocite{*}
\bibliography{ref}

\end{document}